%% file: main.tex
\pdfoutput=1
\documentclass[11pt]{article}
\usepackage{acl}
\usepackage{times} 
\usepackage{latexsym}
\usepackage[T1]{fontenc}
\usepackage[utf8]{inputenc}
\usepackage{microtype}
\usepackage{inconsolata}
\usepackage{graphicx}
\usepackage{amssymb}

\input{setup}

\AtBeginDocument{%
  \providecommand\BibTeX{{%
    \normalfont B\kern-0.5em{\scshape i\kern-0.25em b}\kern-0.8em\TeX}}}

\begin{document}

\title{\ourtool: Static Program Slicing Using Language Models With Dataflow-Aware Pretraining and Constrained Decoding} 

\author{
  Pengfei He$^{1}$, Shaowei Wang$^{1}$, Tse-Hsun Chen$^{2}$, Muhammad Asaduzzaman$^{3}$\\
  $^{1}$ Department of Computer Science, University of Manitoba, Winnipeg, Canada\\
  $^{2}$ School of Engineering and Computer Science, Concordia University, Montreal, Canada\\
  $^{3}$   School of Computer Science, University of Windsor, Windsor, Canada\\
  \texttt{hep2@myumanitoba.ca, shaowei.wang@umanitoba.ca, peterc@encs.concordia.ca, masaduzz@uwindsor.ca}
}



\maketitle
\input{0-abstract}

\input{1-introduction}

\input{3-challenges}

\input{4-method}

\input{5-experiment}

\input{6-results}

\input{conclusion}

\clearpage
\input{limitation}

\newpage

\bibliography{main}

\newpage

\appendix

\input{appendix}

\end{document}

%% file: setup.tex
\usepackage{algorithmic}
\usepackage{textcomp}
\usepackage{xspace}
\usepackage[ruled, lined, linesnumbered, commentsnumbered, longend]{algorithm2e}
\usepackage{multirow,multicol}
\usepackage{colortbl}
\usepackage{caption}
\usepackage{subfigure}
\usepackage{tcolorbox}
\usepackage{mathtools}
\usepackage{tikz}
\usepackage{units}
\usepackage{listings}
\usepackage{ulem}
\usepackage{bbding}
\usepackage{bm}
\definecolor{backcolour}{rgb}{0.95,0.95,0.92}
\usepackage{enumitem}

\SetCommentSty{mycommfont}

\newcommand*\circled[1]{\tikz[baseline=(char.base)]{
            \node[shape=circle,fill,inner sep=1pt,font=\footnotesize] (char) {\textcolor{white}{#1}};}}

\usepackage{tabularx}
\usepackage{listings}
\usepackage{color}

\definecolor{dkgreen}{rgb}{0,0.6,0}
\definecolor{gray}{rgb}{0.5,0.5,0.5}
\definecolor{mauve}{rgb}{0.58,0,0.82}

\usepackage{xcolor}

\graphicspath{ {./figures/} }
\usepackage{textgreek}
\usepackage{float}

\newcommand{\ourtool}{\textsc{Sliceformer}\xspace}

%% file: 0-abstract.tex
\begin{abstract}

Static program slicing is a fundamental software engineering technique for isolating code relevant to specific variables. While recent learning-based approaches using language models (LMs) show promise in automating slice prediction, they suffer from \textit{inaccurate dependency modeling} and \textit{unconstrained generation}, where LMs fail to capture precise data flow relations and produce slices containing hallucinated tokens and statements.
To address these challenges, we propose \ourtool, a novel approach that reformulates static program slicing as a sequence-to-sequence task using small language models such as CodeT5+. \ourtool introduces two key innovations that directly target the identified limitations. First, to improve dependency modeling, we design \textit{dataflow-aware pretraining objectives} that leverage data flow graphs ($\mathcal{DFG}$) to teach models data dependencies through dataflow-preserving statement permutation and dataflow-aware span corruption. Second, to eliminate hallucination, we develop a \textit{constrained decoding} mechanism that enforces both lexical and syntactic constraints. We evaluate \ourtool on Java and Python program slicing benchmarks, demonstrating consistent improvements over state-of-the-art baselines with up to 22\% gain in ExactMatch. 
\end{abstract}

%% file: 1-introduction.tex
\section{Introduction}\label{sec:intro}

Static program slicing identifies code relevant to a given slicing criterion and has proven essential for vulnerability analysis~\citep{vdpm,vdp} and debugging~\citep{slice1984,xu2005brief,slicevd}. Unlike dynamic slicing, static slicing operates on the code dependence graph constructed via static analysis, without requiring program execution~\citep{harman2001overview}, thereby offering broader practicality~\citep{acharya2011practical,xu2005brief}.

Recent work has explored learning-based approaches for program slicing using LMs~\citep{learning,llmslicer}. 
However, the learning-based approaches face significant hurdles, as demonstrated by our empirical investigation (Section~\ref{sec:motivation}):  \circled{1}~\textbf{Inaccurate dependency identification}~\citep{learning}: general LLMs or directly fine-tuned LMs fail to precisely capture data dependencies, frequently missing relevant statements or including extraneous ones. \circled{2}~\textbf{Unconstrained generation}: employing large proprietary LMs with prompting techniques (e.g. CoT) suffers from hallucination~\citep{llmslicer}, where models generate hallucinated tokens or statements not in the original code, violating the requirement that slices must be exact subsequences of the input.

To address these challenges, we propose \ourtool, which enhances standard supervised fine-tuning (SFT) with two key contributions: \textbf{dataflow-aware pretraining} and \textbf{constrained decoding}. The former equips the model with dataflow semantic understanding before SFT, while the latter operates at inference time to prevent hallucinated tokens and invalid statement generation.

\noindent\textbf{Dataflow-aware pretraining}: We introduce two novel pretraining objectives based on Data Flow Graph ($\mathcal{DFG}$) structures~\citep{graphcodebert}: (1)~\textit{Dataflow-preserving statement permutation}, which trains the model to identify statement dependencies by generating valid code permutations that respect dataflow constraints; (2)~\textit{Dataflow-aware span corruption}, which employs dataflow-guided masking to force the model to reconstruct code based on dependency relationships rather than superficial patterns.

\noindent\textbf{Constrained decoding}: We propose a decoding mechanism that enforces two constraints:
(1)~\textbf{Lexical constraint}, which restricts the vocabulary to tokens appearing in the original code; and
(2)~\textbf{Syntactic constraint}, which leverages the monotonicity of Tree Similarity Edit Distance (TSED)~\citep{tsed} to filter AST-invalid candidates. These constraints mitigate hallucinated tokens and invalid statement generation.

We evaluate \ourtool on Java and Python program slicing tasks. \ourtool consistently outperforms state-of-the-art baselines~\citep{llmslicer,learning} across all evaluation metrics, achieving ExactMatch improvements of 6.4\% and 21.9\%, respectively. 

In summary, we make the following contributions:

\begin{itemize}
    \item We propose \ourtool, an approach that integrates dataflow-aware pre-training objectives and constrained decoding to explicitly capture data relations for accurate program slicing.
    \item To the best of our knowledge, we are the first to introduce novel dataflow-aware statement permutation and span corruption objectives for pre-training, and TSED-monotonicity constraint for decoding.
    \item Through extensive evaluation on Java and Python benchmarks, \ourtool consistently outperforms state-of-the-art static and learning-based slicing baselines.
    \item We publicly release our code and datasets.\footnote{https://anonymous.4open.science/r/staticsliceT5-4E22}
\end{itemize}

%% file: 3-challenges.tex
\section{Problem Formulation and Challenges}\label{sec:overview}
We formulate program slicing as a sequence-to-sequence learning problem and identify key limitations that motivate our approach.

\subsection{Problem Formulation}\label{sec: problem}
\input{case}

We formulate \textbf{static program slicing} as a sequence-to-sequence learning task. The input is:
\begin{align}
\bm{x} = \{s_1, s_2, \dots, s_N; v; [n]\}
\end{align}
where $s_i$ is the $i$-th statement, $v$ is the slicing criterion (variable of interest), and $[n]$ is the line number where $v$ appears. Given $\bm{x}$, a slicer LM $P(\bm{y}|\bm{x})$ to predict the \textit{backward program slice}~\citep{slice1984} $\bm{y}$:
\begin{align}
    \bm{y} = \{s_{i_1}, s_{i_2}, \dots, s_{i_k}\} 
    &\subseteq \{s_1, s_2, \dots, s_{n}\}, \notag \\
    &\quad i_1 < i_2 < \dots < i_k
\end{align}
where $\bm{y} \subseteq \bm{x}$ comprises the statements that semantically influence $v$. The output must satisfy:
\begin{itemize}
  \item \textbf{Accuracy}: $\bm{y}$ includes all and only relevant statements, with no extraneous or missing statements.
  \item \textbf{Element preservation}: $\bm{y}$ is an exact subsequence of $\bm{x}$: every token and statement must be precisely extracted from $\bm{x}$ without modification or hallucination.
\end{itemize}

\subsection{Related Work and Limitations}\label{sec:motivation}

Traditional static slicing tools, such as JavaSlicer~\citep{javaslicer} and CPP-Slicer~\citep{cppslicer}, rely on static dependency analysis to construct System Dependence Graphs (SDGs) and compute slices via graph reachability~\citep{javaslicerpaper}.

Recently, learning-based approaches have attracted growing interest~\citep{learning,llmslicer}.
\citet{learning} predict dependencies between slicing criteria and program statements via binary classification using CodeBERT~\citep{codebert} and GraphCodeBERT~\citep{graphcodebert} embeddings. Although GraphCodeBERT incorporates dataflow-aware pretraining, it is an embedding model with objectives that differ from ours. Moreover, NS-Slicer is not end-to-end and operates on isolated statements, which restricts the available context to partial functions and consequently limits slicing performance. \citet{llmslicer} investigated LLMs (e.g., GPT-4o~\citep{gpt4}, GPT-3.5-Turbo, and Gemma~\citep{gemma2}) combined with advanced prompting techniques such as Retrieval-Augmented Generation (RAG) and Chain-of-Thought (CoT). However, proprietary LLMs are computationally expensive and prone to severe hallucinations~\citep{hallucination1,hallucination2}, which limits their reliability for program analysis tasks.

As a result, existing learning-based slicing methods suffer from the following two key challenges:


\begin{itemize}[leftmargin=0.in]
    \item[] \circled{1} \textbf{Inaccurate dependency identification.} LMs struggle to precisely capture data dependencies, often omitting relevant statements or including irrelevant ones. For instance, in Figure~\ref{fig:motivation} (Example 1), the generated slice incorrectly includes lines 9--10 (\texttt{temp = A; A = C;}), which are unrelated to the slicing criterion. Instead of identifying true dependencies, the model relies on surface-level patterns or positional proximity, violating the \textit{accuracy} property.

    \item[] \circled{2} \textbf{Unconstrained generation.} LMs hallucinate elements absent from the original program, violating \textit{element preservation}. We observe: \textbf{(a) Token-level:} original identifiers are replaced with hallucinated ones (e.g., \texttt{keta} instead of \texttt{codepoint} in Example 2), leading to syntactic errors~\citep{errors}; \textbf{(b) Statement-level:} spurious logic or repetitions are introduced (e.g., \texttt{y * 10 * y * y * y * z * ten ...} in Example 3 includes incorrect sub-expressions such as \texttt{z * ten * 10 * y}).
\end{itemize}

%% file: case.tex
\begin{figure}[!ht]
\centering

\begin{minipage}[t]{0.48\linewidth}
\begin{tcolorbox}[
  boxsep=1pt,
  left=1pt,right=1pt,top=1pt,bottom=1pt,
  height=5.2cm,
  valign=top
]
\begin{lstlisting}[language=Java,frame=single,framerule=0pt,basicstyle=\ttfamily\scriptsize]
// Example (1) Inaccurate dependency identification
// Expected:
7: int temp
8: if(C <= A){
12: temp = B;

// Generated:
7: int temp
8: if(C <= A){
9: (*@\textcolor{red}{temp = A;}@*)
10:(*@\textcolor{red}{A = C;}@*)
12: temp = B;
\end{lstlisting}
\end{tcolorbox}
\end{minipage}
\hfill
\begin{minipage}[t]{0.48\linewidth}
\begin{tcolorbox}[
  boxsep=1pt,
  left=1pt,right=1pt,top=1pt,bottom=1pt,
  height=5.2cm,
  valign=top
]
\begin{lstlisting}[language=Java,frame=single,framerule=0pt,basicstyle=\ttfamily\scriptsize]
// Example (2) Non-existent variable replacement
// Expected:
7: int cnt = 0;
10: for(int i=cnt;i>=0;i--) {
11: if(i>0) {long y = x[i];
12: long Codepoint = 97+y};

// Generated:
7: int cnt = 0;
10: for(int i=cnt;i>=0;i--) {
11: if(i>0) {long y = x[i];
12: long (*@\textcolor{red}{keta}@*) = 97+y};
\end{lstlisting}
\end{tcolorbox}
\end{minipage}

\vspace{3pt}

\begin{tcolorbox}[boxsep=1pt,left=1pt,right=1pt,top=1pt,bottom=1pt]
\begin{lstlisting}[language=Java,frame=single,framerule=0pt,basicstyle=\ttfamily\scriptsize]
// Example (3) Hallucinated statement generation
// Expected:
4: int one = 0, five = 0, ten = n;
5: try {
6: if (one * 1 + five * 5 + ten * 10 > y)

// Generated:
4: int one = 0, five = 0, ten = n;
5: try {
6: if (one * 1 + five * 5 + ten * 10 > y
   (*@\textcolor{red}{* 10 * y * y * y * z * ten * 10 * y * z * ten *}@*)
\end{lstlisting}
\end{tcolorbox}

\caption{Motivating examples of incorrect program slices produced by LMs.}

\vspace{-0.7cm}
\label{fig:motivation}
\end{figure}

%% file: 4-method.tex
\section{Methodology}\label{sec:framework}

\begin{figure}
    \centering
    \includegraphics[width=0.9\linewidth]{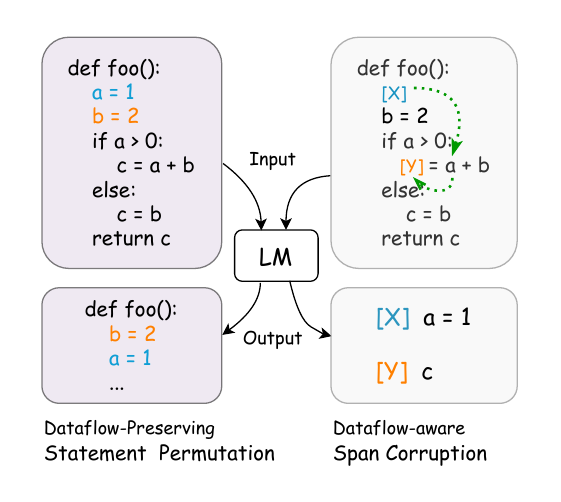}
    \caption{Illustration of dataflow-aware pre-training objectives. Left: given a snippet, we generate a statement permutation that preserves the dataflow of the original program. Right: we select a variable $a$ with a dataflow chain $[X] \rightarrow a \rightarrow [Y]$, then mask its parent $[X]$ (where it comes from) and child $[Y]$ (where it flows to). The pre-training objective is to recover both $[X]$ and $[Y]$.} 
    \label{fig:pretraining}
    \vspace{-0.7cm}
\end{figure}

    
We introduce novel \textbf{dataflow-aware pre-training} and \textbf{lexical–syntactic constrained decoding} to enhance SFT-based LMs for accurate, element-preserving program slicing.

\noindent\textbf{Dataflow-Aware Pretraining.} To tackle Challenge~\circled{1}, we introduce a pretraining strategy that explicitly models program dependencies via two objectives. First, \textit{dataflow-preserving statement permutation} (Section~\ref{sec:perm}) trains the model to recognize statement dependencies by generating valid permutations that respect dataflow constraints. Second, \textit{graph-aware span corruption} (Section~\ref{sec:span}) applies $\mathcal{DFG}$-guided masking to force code reconstruction based on dependency relationships. We then perform SFT on labeled slicing data to specialize the model for slicing.

\noindent\textbf{Constrained Decoding.} To address Challenge~\circled{2}, we design a training-free constrained decoding mechanism (Section~\ref{sec:decoding}) that enforces lexical and syntactic correctness during inference. We apply two complementary constraints: \textbf{(1) lexical constraints}, which restrict generation to tokens appearing in the input code, and \textbf{(2) syntactic constraints}, which evaluate structural coherence at statement boundaries using Tree Similarity of Edit Distance (TSED)~\citep{tsed}. Since valid slices are subsequences of the input, TSED should increase monotonically. When the TSED score decreases, indicating structural errors, we terminate that generation path. These constraints ensure extractive, structurally sound slices.

\subsection{Dataflow-Aware Pretraining}\label{sec:pretraining}

\begin{algorithm}[!ht]
\caption{Dataflow-Preserving Statement Permutation}
\label{alg:perm}

\footnotesize
\SetKwInOut{KwIn}{Input}
\SetKwInOut{KwOut}{Output}
\SetKwFunction{ExtractVars}{ExtractVars}
\SetKwFunction{SameBasicBlock}{SameBasicBlock}

\KwIn{Code snippet $C = \{s_1, s_2, \ldots, s_n\}$, $\mathcal{DFG} = (V, E)$}
\KwOut{Permuted code snippet $C'$}
 
$\mathcal{P} \leftarrow \emptyset$

\ForEach{pair $(s_i, s_j)$ where $i < j$}{
    \tcp{Check independence: same basic block and no data edge}
    $hasEdge \leftarrow \text{false}$

    \ForEach{$v_a \in s_i, v_b \in s_j$}{
        \If{$\langle v_a, v_b \rangle \in E$ \textbf{or} $\langle v_b, v_a \rangle \in E$}{
            $hasEdge \leftarrow \text{true}$; \textbf{break}
        }
    }

    \If{$\SameBasicBlock(s_i, s_j)$ \textbf{and} $\neg hasEdge$}{
        $\mathcal{P} \leftarrow \mathcal{P} \cup \{(s_i, s_j)\}$
    }
}

\tcp{Apply random permutation if possible}
\eIf{$\mathcal{P} \neq \emptyset$}{
    $(s_i, s_j) \leftarrow$ random pair from $\mathcal{P}$\\
    $C' \leftarrow $ swap $s_i$ and $s_j$ in $C$
}{
    $C' \leftarrow C$
}

\KwRet $C'$
\end{algorithm}

\begin{algorithm}[!htb]
\caption{Dataflow-Aware Span Corruption}
\label{alg:span}

\footnotesize
\SetKwInOut{KwIn}{Input}
\SetKwInOut{KwOut}{Output}
\SetKwFunction{Size}{Size}
\SetKwFunction{Mask}{Mask}
\SetKwFunction{GetStmt}{GetStmt}
\SetKwFunction{Random}{Random}

\KwIn{Code snippet $C$, $\mathcal{DFG} = (V, E)$, Mask ratio $r$}
\KwOut{Masked code snippet $C_{\text{masked}}$}

$m \leftarrow r \times \Size(C)$ \tcp{Target masked tokens}
$masked \leftarrow 0$

\tcp{Iteratively mask}
\While{$masked < m$ \textbf{and} $V \neq \emptyset$}{
    \tcp{Randomly select a variable from $\mathcal{DFG}$}
    $v \leftarrow \Random(V)$

    \tcp{Find parents (where v comes from/ who defines v)}
    $Parents \leftarrow \{u \mid \langle u, v \rangle \in E\}$

    \tcp{Find children (where v goes / who uses v)}
    $Children \leftarrow \{w \mid \langle v, w \rangle \in E\}$

    \tcp{Randomly decide masking granularity}
    \eIf{\Random() $< 0.5$}{
        \tcp{Fine-grained: mask variables}
        $Mask(Parents \cup Children)$
    }{
        \tcp{Coarse-grained: mask statements}
        $S_{parents} \leftarrow \{\GetStmt(u) \mid u \in Parents\}$

        $S_{children} \leftarrow \{\GetStmt(w) \mid w \in Children\}$

        $Mask(S_{parents} \cup S_{children})$
    }

    $masked \leftarrow  masked + \Size(\mathcal{M})$
}
\KwRet{$C_{\text{masked}}\leftarrow C$}
\end{algorithm}

To enable the model to better understand data relations crucial for program slicing, we introduce two novel dataflow-aware pretraining objectives. Figure~\ref{fig:pretraining} illustrates our dataflow-aware pretraining objectives.

\textbf{Data Flow Graph ($\mathcal{DFG}$).}
Given source code $C$, $\mathcal{DFG}= (V, E), \text{where } V = \{v_1,\ldots, v_k\}$: The variable sequence extracted from AST terminals (leaves). and Edges $E = \{\varepsilon_1, \ldots, \varepsilon_l\}$: A directed edge $\varepsilon = \langle v_i, v_j \rangle$ indicates that there is a data dependency between the two variables~\citep{graphcodebert}. $\mathcal{DFG}$s are the core inherent structure for determining program slices~\citep{javaslicerpaper}.

\subsubsection{Dataflow-Preserving Statement Permutation}\label{sec:perm}
Conventional permutation-based pretraining methods~\citep{bart} randomly permute sentences and train models to recover the original order for capturing their semantic dependency. In code, however, legal orderings are not unique: a given $\mathcal{DFG}$ permits multiple valid statement orders. Statements without data dependencies within the same basic block can be reordered without affecting program semantics. As illustrated in Figure~\ref{fig:pretraining}, the two independent statements in a basic block are equivalent under any ordering.

To improve the model's capability to recognize the true dependencies, we introduce a dataflow-preserving statement permutation pretraining objective. Given a code snippet, we train the model to generate equivalent variants while preserving dataflow invariance. The training labels are obtained by reordering independent statements within each basic block as shown in Algorithm~\ref{alg:perm}. This forces the model to learn which statements are data-independent, a fundamental capability for identifying relevant statements in program slicing.

\subsubsection{Dataflow-Aware Span Corruption}\label{sec:span}

Span corruption is a well-established pre-training objective for language models, in which consecutive tokens are masked and replaced with sentinel tokens~\citep{t5,bart}. AST-T5~\citep{astt5} extends this paradigm by incorporating syntactic structure, masking subtrees derived from abstract syntax trees (ASTs) to improve LM's ability to comprehend ASTs . Compared to ASTs, data-flow graphs are structurally simpler and avoid unnecessarily deep hierarchies~\citep{graphcodebert}. As a result, $\mathcal{DFG}$s are well-suited for program slicing, where capturing long-range data dependencies is crucial. Therefore, we design pre-training objective based on $\mathcal{DFG}$s to improve LM's ability to understand data-flow. 

\noindent\textbf{Two-granularity dataflow-guided masking.}
A core design principle of our approach is that, \emph{given a node in the $\mathcal{DFG}$, the model is trained to reconstruct where the value comes from and where it flows to}.
Accordingly, each masked span corresponds to semantically coherent code elements connected by def--use relations.
To increase the diversity of reconstruction objectives, we perform masking at two different granularities (Algorithm~\ref{alg:span}):

For example, in Figure~2, given the variable node \texttt{a} in the statement \texttt{c = a + b}, the value of \texttt{a} is defined in the statement \texttt{a = 1} and subsequently flows into \texttt{c}.
In this case, we apply \emph{coarse-grained masking} to the defining statement \texttt{a = 1}, and \emph{fine-grained masking} to the variable usage \texttt{c} in \texttt{c = a + b}. This forces the model to reconstruct the dataflow chain associated with \texttt{a}, explicitly learning variable-level data dependencies.

By combining these two granularities, our approach enables the model to capture both fine-grained variable-level dependencies and coarse-grained statement-level dataflow, while ensuring that every masked unit remains semantically coherent within the DFG.

\subsubsection{Supervised Fine-Tuning}
Our dataflow-aware pretraining objectives endow the model with a strong understanding of data dependencies. We then perform supervised fine-tuning (SFT) on labeled slicing data to directly train the model for program slicing.
To encourage structured, task-specific outputs, we augment the input–output format with special control markers, following prior work on conditional text generation~\citep{condition,condition2,grammert5}. Specifically, we introduce markers to denote line numbers, code, slicing criteria, and slices (e.g., \textit{<code>}, \textit{<criterion>}, \textit{<slice>}), which provide explicit structural cues and guide the model to generate well-formed outputs.
\begin{algorithm}[h]
\caption{Constrained Beam Search with Lexical Constraint and Syntactic Constraint}
\label{alg:constrained_beam}

\footnotesize
\SetKwInOut{KwIn}{Input}
\SetKwInOut{KwOut}{Output}
\SetKwFunction{GetAllowedTokens}{GetAllowedTokens}
\SetKwFunction{ApplyMask}{ApplyMask}
\SetKwFunction{NextTokenScores}{NextTokenScores}
\SetKwFunction{TopK}{TopK}
\SetKwFunction{TSED}{TSED}
\SetKwFunction{IsEOS}{IsEOS}
\SetKwFunction{IsStatementComplete}{IsStatementComplete}

\KwIn{Input $\bm{x}$; LM $\mathcal{M}$; Vocabulary $\mathcal{V}$; Beam size $K$; Max output length $L$}
\KwOut{Program slice $\mathcal{Y}$}

\tcp{Determine lexically allowed tokens}
$\mathcal{A} \leftarrow \GetAllowedTokens(\bm{x})$

\tcp{Initialize beam}
$\mathcal{B} \leftarrow \{(\bm{y} = [], s = 0, t_{stmt} = 0)\}$

\For{$t = 1$ \KwTo $L$}{
    $\mathcal{B}_{\text{next}} \leftarrow \emptyset$

    \ForEach{beam $(\bm{y}, s, t_{stmt}) \in \mathcal{B}$}{
        \tcp{Apply Lexical Constraint}
        $mask \leftarrow \ApplyMask(\mathcal{A}, \mathcal{V})$\\
        $\bm{p} \leftarrow \NextTokenScores(\mathcal{M}, \bm{y}, mask)$\\
        $\{(z_k, p_k)\}_{k=1}^K \leftarrow \TopK(\bm{p}, K)$

        \tcp{Expand each beam}
        \For{$k = 1$ \KwTo $K$}{
            $\bm{y}' \leftarrow \bm{y} \mathbin{\Vert} z_k$\\
            $s' \leftarrow s + \log(p_k)$\\
            $t'_{stmt} \leftarrow t_{stmt}$

            \tcp{Apply Syntactic Constraint at statement boundary}
            \If{\IsStatementComplete($\bm{y}'$)}{
                $t_{\text{cur}} \leftarrow \TSED(\bm{x}, \bm{y}')$\\

                \tcp{Skip if TSED decreases (syntactic error)}
                \If{$t_{\text{cur}} < t_{stmt}$}{
                    \textbf{continue}
                }
                $t'_{stmt} \leftarrow t_{\text{cur}}$
            }

            \tcp{Skip if end of sequence}
            \If{\IsEOS($z_k$)}{
                \textbf{continue}
            }

            $\mathcal{B}_{\text{next}} \leftarrow \mathcal{B}_{\text{next}} \cup \{(\bm{y}', s', t'_{stmt})\}$
        }
    }
    $\mathcal{B} \leftarrow \TopK(\mathcal{B}_{\text{next}}, K)$
}
\tcp{Return the top-1 output sequence}
\KwRet{$\mathcal{Y} \leftarrow \arg\max_{(\bm{y}, s, t_{stmt}) \in \mathcal{B}} s$}
\label{alg:beam-lex-tsed}
\end{algorithm}

\subsection{Constrained Decoding}\label{sec:decoding}
While our dataflow-aware pretraining provides the model with a strong understanding of code dependencies, it does not guarantee that the generated slices will be free from hallucinated tokens or structural errors. To address this, we introduce a constrained decoding mechanism. 

Our approach modifies the standard beam search by applying two types of constraints at different granularities, as shown in Algorithm~\ref{alg:constrained_beam} and Figure~\ref{fig:decoding}. At each token generation step, we apply a \textit{lexical constraint} to restrict the vocabulary to only tokens appearing in the original code snippet. Also, when a complete statement is generated, we apply a \textit{syntactic constraint} at the statement boundary to filter out candidates that violate structural coherence. 

\subsubsection{Lexical Constraint}

A challenge in applying LMs to program slicing is their tendency to produce \textit{hallucinated} tokens: identifiers, keywords, or operators absent from the input code (Figure~\ref{fig:motivation}). This occurs because LMs decode from an unconstrained vocabulary. For example, CodeT5+ samples from a fixed vocabulary of 32,100 tokens~\citep{codet5+} without awareness of which tokens appear in the input.


To address this issue, we introduce a \textit{lexical constraint} that restricts decoding to tokens present in the input sequence. Specifically, any token absent from the original code snippet $\bm{x}$ is assigned a logit of $-\infty$ before softmax, ensuring zero probability. This hard constraint prevents hallucinated identifiers and enforces strictly extractive token generation, producing slices that are lexically faithful to the original code snippet.

\subsubsection{Syntactic Constraint}

\begin{figure}
  \centering
    {\includegraphics[width=0.9\linewidth]{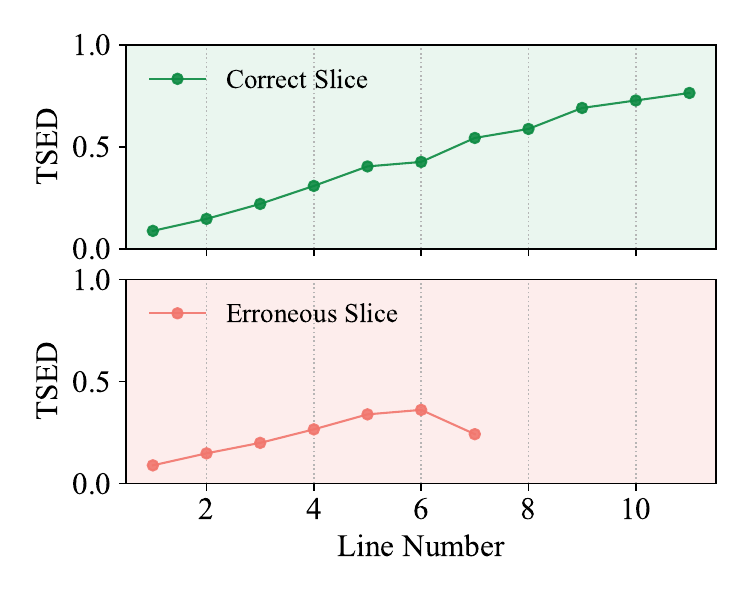}}
    \caption{TSED scores at statement boundaries for syntactically correct versus erroneous program slices based on the same input code snippet. The scores are derived from Example (3) in Figure \ref{fig:motivation}. When a syntactic error is introduced into the generated slice (Line 6), the TSED score deviates from its expected monotonic increasing pattern and instead decreases at the statement boundary.}\label{fig:mono}
    \vspace{-0.3cm}
\end{figure}

While the lexical constraint restricts generation to input tokens, it is \textit{order-agnostic} and does not ensure structural coherence. As a result, the model may still produce syntactically invalid slices, such as \textit{over-generation}~\citep{over}, where tokens, though individually valid, are repeated or misaligned. For example, Example (3) in Figure~\ref{fig:motivation} shows a slice that satisfies the lexical constraint but contains a structural error due to incorrect ordering. 

To address this limitation, we introduce a \textit{syntactic constraint} based on the Tree Similarity of Edit Distance (TSED)~\citep{tsed}, a similarity metric that quantifies the syntactic distance between two code snippets by comparing their abstract syntax trees (ASTs). TSED is defined as:
\begin{align}
TSED(T_x, T_y) = 1 - \frac{\min_{\text{ops}} \sum_{i=1}^{n} w(op_i)}{\max(\text{Nodes}(T_x, T_y))}\label{fml:tsed}
\end{align}%
\noindent where $\textit{ops}$ denotes the sequence of $n$ edit operations (insertions, deletions, substitutions) required to transform the AST $T_x$ of the original code snippet into the AST $T_y$ of the (partial) generated slice, $w(op_i)$ represents the cost of the $i$-th operation, and the distance is normalized by the maximum number of nodes in the two trees to account for variations in code size and complexity.

\noindent\textbf{TSED monotonicity} The key insight underlying syntactic constraint is that, \emph{since every valid program slice is a \textit{subsequence} of the input code, the autoregressive generation process should produce outputs that progressively resemble the original code's AST.} Consequently, the TSED score between the input and the generated slice should \textit{increase monotonically} at statement boundaries as more correct statements are appended. As illustrated in Figure~\ref{fig:mono}, when no structural error occurs, the TSED score exhibits a monotonic increasing pattern at each statement boundary. However, when syntactic errors occur, such as incorrectly ordered statements, they disrupt the structural coherence and cause the TSED score to \textit{decrease}. To detect and reject malformed candidates, we terminate a beam path early and discard its output whenever the TSED score drops at a statement boundary.

%% file: 5-experiment.tex
\section{Experimental Setting}\label{sec:experimentalsetting}


\subsection{Datasets and Metrics}
We use the Python and Java subsets of the CodeNet-Slice dataset~\citep{learning} using the training split, and evaluate \ourtool on the test split. Ground-truth slices are obtained using language-specific slicing tools: JavaSlicer~\citep{javaslicer} for Java and a Python slicer adapted from the JavaSlicer implementation~\citep{javaslicerpaper}. Detailed dataset statistics are reported in Table~\ref{tab:dataset}.

Following previous studies~\citep{learning,llmslicer}, we evaluate the performance of \ourtool using four metrics: 1) \textbf{Dependence Accuracy (Acc-D)} is defined as the average ratio of correctly predicted dependencies to the total number of ground-truth dependencies; 2) \textbf{Exact Match} calculates the percentage of instances where the generated program slice exactly matches the ground-truth slice; 3) \textbf{CodeBLEU} is a composite text similarity metric specifically designed for coding tasks; 4) \textbf{TSED} is a code similarity metric that compares the ASTs of the generated and reference code by measuring the minimum edit operations required to transform one tree into another. 

\subsection{Baselines}\label{baselines}
\textbf{LLM-based slicers~\citep{llmslicer}} leverage proprietary LLMs to perform program slicing using prompt engineering techniques, Zero-shot, Retrieval-Augmented Generation (RAG)~\citep{allyouneed}, and Chain-of-Thought (CoT)~\citep{cot}. We select two state-of-the-art foundation language models, GPT-4o-mini~\citep{gpt4} and GPT-5~\citep{openai2025gpt5}, as the base LLMs.

\noindent\textbf{NS-slicer~\citep{learning}} formulates the program slicing task as a binary classification problem. It applies CodeBERT~\citep{codebert} and GraphCodeBERT~\citep{graphcodebert} to learn statement embeddings and then computes the distance to make predictions.

\noindent\textbf{Fine-tuned models.} We compare against models obtained by directly applying SFT to LMs without dataflow-aware pretraining and constrained decoding). Specifically, we fine-tune CodeLlama-7B~\citep{codellama}, Qwen3-8B~\citep{qwen3}, and CodeT5+~\citep{codet5+} on the training set. For CodeLlama and Qwen3, we employ QLoRA~\citep{qlora} to reduce hardware requirements.

\subsection{Implementation Details}

\noindent\textbf{Pretraining} We implement \ourtool on top of CodeT5+~\citep{codet5+} (0.7B), a lightweight model that has been shown to be effective for a wide range of coding tasks~\citep{codet5+,li2024few,jiao2023evaluation,yin2024rectifier}.
The model is pretrained on the Python and Java subsets of the CodeSearchNet dataset~\citep{codesearchnet}, comprising approximately 1.0M functions. Following GraphCodeBERT~\citep{graphcodebert}, each function is first parsed into an AST using Tree-Sitter~\citep{tree-sitter} to identify variables, after which a dataflow graph is constructed by following GraphCodeBERT's implementation~\citep{graphcodebert}.
For span corruption, we mask 25\% of tokens, following the empirical setting in~\citep{astt5}. For statement permutation, we randomly permute up to three statements per code sample. Pretraining is performed with a context length of 512 tokens and a batch size of 32 for 100K steps. All experiments are conducted on four NVIDIA RTX 3090 GPUs (24GB each).

\noindent\textbf{Fine-tuning} We set the input and output lengths to 512 tokens. The model is trained using the AdamW optimizer with a batch size of 32, a learning rate of $5 \times 10^{-5}$, and 1,000 warmup steps over 10 epochs. All other settings follow the default CodeT5+ configuration. 
We perform supervised fine-tuning on the training split of the Python and Java subsets of the CodeNet-Slice dataset~\citep{learning}.

\noindent\textbf{Constrained Decoding}
We use a beam size of 3. For TSED, we adopt the implementation from the original paper, which supports both Java and Python. The lexical constraints are implemented in a HuggingFace–compatible manner, following the LogitsProcessor and Constraint interfaces~\citep{hfgeneration}.

%% file: 6-results.tex
\section{Results}\label{sec:results}
\input{rq1}

\input{rq2}

\input{rq3}

%% file: rq1.tex
\subsection{Effectiveness of \ourtool}
\begin{table*}[!htbp]
\centering
\scriptsize
\renewcommand{\arraystretch}{1.2}
\begin{tabular}{l|cccc|cccc}
\hline
\multicolumn{1}{c|}{\multirow{2}{*}{\textbf{Methods}}} & \multicolumn{4}{c|}{\textbf{Java}}                                                            & \multicolumn{4}{c}{\textbf{Python}}                                                          \\
\multicolumn{1}{c|}{}                                  & \textbf{Acc-D} & \textbf{ExactMatch} & \multicolumn{1}{l}{\textbf{CodeBLEU}} & \textbf{TSED}  & \textbf{Acc-D} & \textbf{ExactMatch} & \textbf{CodeBLEU} & \multicolumn{1}{l}{\textbf{TSED}} \\ \hline
LLM-based slicer (GPT-4/Zero-shot)                                      & 14.76          & 0.00                & 20.10                                 & 35.53          & 12.34          & 0.00                & 18.45             & 32.18                             \\
LLM-based slicer (GPT-4/RAG)                                            & 51.70          & 0.00                & 65.18                                 & 59.93          & 48.22          & 0.00                & 61.34             & 56.71                             \\
LLM-based slicer (GPT-4/COT)                                            & 56.84          & 0.00                & 68.41                                 & 59.68          & 53.17          & 0.00                & 65.23             & 57.42                             \\
LLM-based slicer (GPT-5/Zero-shot)                                      & 18.42          & 0.00                & 24.67                                 & 39.21          & 15.78          & 0.00                & 22.13             & 36.54                             \\
LLM-based slicer (GPT-5/RAG)                                            & 55.13          & 7.00                & 68.92                                 & 62.45          & 51.68          & 11.00                & 64.77             & 59.33                             \\
LLM-based slicer (GPT-5/COT)                                            & 60.27          & 14.00                & 71.35                                 & 63.81          & 56.94          & 13.00                & 68.56             & 61.27                             \\
\hline
NS-slicer (CodeBERT)                                   & 95.65          & 81.72               & 88.41                                 & 91.00          & 82.47          & 56.32               & 74.68             & 78.91                             \\
NS-slicer (GraphBERT)                                  & 96.51          & 85.77               & 89.26                                 & 90.35          & 84.92          & 61.25               & 76.84             & 80.12                             \\
\hline
CodeLlama (SFT)                                     & 82.83          & 75.27               & 79.10                                 & 81.82          & 75.61          & 68.45               & 72.33             & 74.26                             \\
Qwen3 (SFT)                                            & 87.22          & 80.55               & 80.54                                 & 82.35          & 79.34          & 72.18               & 74.91             & 76.58                             \\
CodeT5+ (SFT)                                          & 95.33          & 87.24               & 89.26                                 & 93.42          & 87.53          & 77.24               & 79.98             & 81.75                             \\
\hline
\ourtool                                & \textbf{98.78} & \textbf{92.20}      & \textbf{93.23}                        & \textbf{97.68} & \textbf{90.85} & \textbf{83.15}      & \textbf{85.35}    & \textbf{89.74}         \\
\hline

\end{tabular}
\caption{Effectiveness comparison among different static learning-based program slicing methods.}
\label{tab:RQ1}
\vspace{-0.4cm}
\end{table*}

\textbf{\ourtool consistently outperforms baselines across two languages and four evaluation metrics.} 
Table~\ref{tab:RQ1} reports the effectiveness of \ourtool compared with existing program slicing methods.
On Java programs, \ourtool attains an ExactMatch score of 92.20\%, significantly outperforming the strongest baseline, NS-slicer (GraphBERT), which achieves 85.77\%. On Python programs, the performance gains are even more pronounced: \ourtool reaches 83.15\% in ExactMatch, compared to only 61.25\% for NS-slicer. Overall, \ourtool surpasses the best-performing baseline by 6.4\% and 21.9\% in ExactMatch on Java and Python, respectively, demonstrating its superior ability to generate fully correct slices across different programming languages.
A qualitative analysis explaining the reasons behind these improvements is provided in Appendix~\ref{sec:qualitativeAnlysis}.

\textbf{\ourtool significantly outperforms fine-tuned vanilla models.} \ourtool consistently exceeds CodeT5+ performance in both languages: in Java, ExactMatch rose from 87.24\% to 92.20\% (5.0\% gain); in Python, improvement was even sharper, rising from 77.24\% to 83.15\% (5.9\% gain). These gains demonstrate that dataflow-aware pretraining combined with constrained decoding (guided by lexical and syntactic knowledge) is critical for achieving high accuracy and faithfulness in program slicing.

\textbf{\ourtool excels at generating fully correct slices} Notably, \ourtool achieves the most significant improvement on the ExactMatch metric, suggesting that it more frequently generates fully correct slices compared to all baselines. In contrast, LLM-based methods yield the weakest performance overall. Although techniques like CoT and RAG lead to improvements compared to the zero-shot setting, the best LLM-based approach (GPT-5 with CoT) only achieves 14.00\% and 13.00\% ExactMatch on Java and Python, respectively, far below the learning-based methods. 


%% file: rq2.tex
\subsection{Ablation Analysis}\label{sec:rq2}

\begin{figure}[!htbp] 
\centering
{\includegraphics[width=1\linewidth]{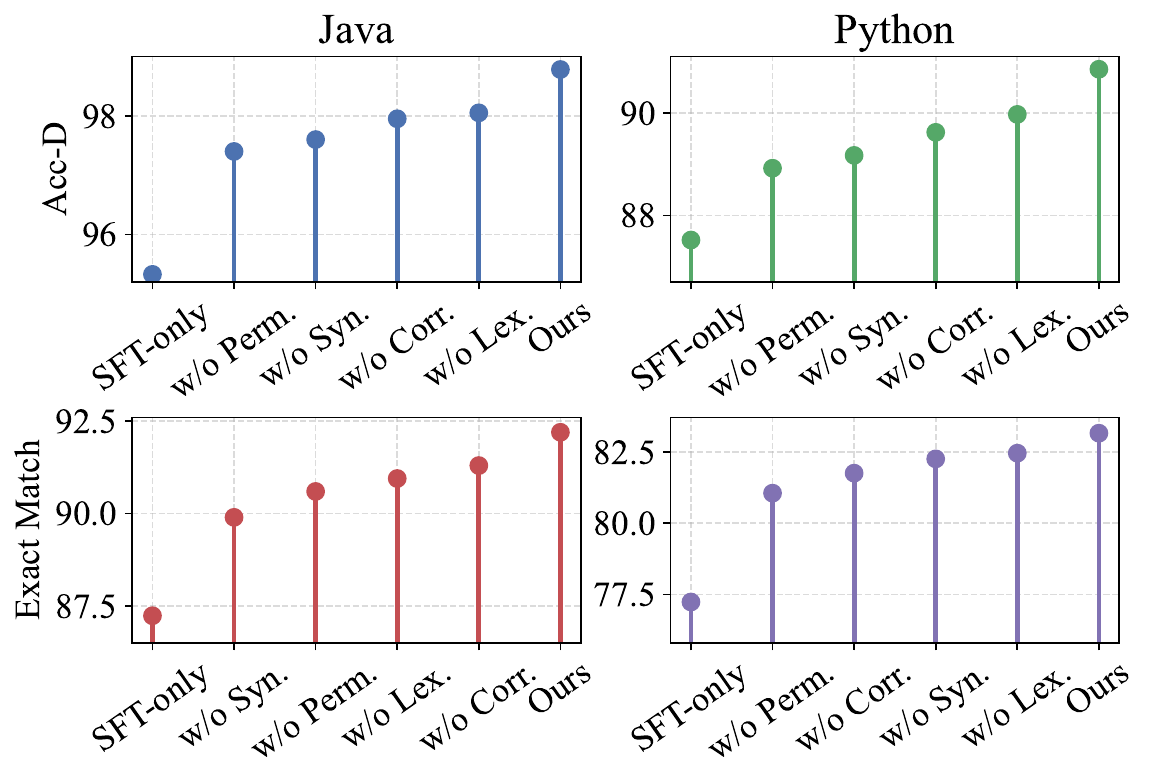}}
       \vspace{-0.5cm}
       \caption{The ablation analysis results highlight the
contribution of each component. A larger performance
drop indicates greater importance. We rank the component importance from
left to right. Abbreviations: Perm. = statement permutation; Corr. = span corruption; Syn. = syntactic constraint; Lex. = lexical constraint.} 
       \label{Fig:RQ2}
\vspace{-0.6cm}
\end{figure}
\textbf{All components contribute to \ourtool's effectiveness.}
Figure~\ref{Fig:RQ2} presents the results of ablation analysis for \ourtool, examining the impact of each component. Overall, both dataflow-aware pretraining strategies (statement permutation and span corruption) and constrained decoding mechanisms (lexical and syntactic constraints) contribute to \ourtool's performance. As shown, the full version of \ourtool consistently outperforms any variant lacking a single component across all evaluation settings.
 
Among these components, span corruption and lexical constraint are the most impactful. The model learns more about data dependencies from dataflow-aware span corruption, which is critical for identifying complete program slices. Meanwhile, lexical constraints ensure that the generated tokens strictly originate from the source code, preventing hallucinations and improving statement-level accuracy. The relatively smaller impact of syntactic constraints can be attributed to the fact that structural errors occur less frequently than lexical errors, as the pretrained model has already learned fundamental code syntax from corpora.

\section{Efficiency of \ourtool}

To evaluate efficiency, we compare inference latency against all baselines.
As shown in Table~\ref{tab:latency}, latency largely depends on the underlying base model.
When running on the same machine, larger models with more parameters consistently incur higher latency.
NS-Slicer, which relies on smaller models, achieves the lowest latency but suffers from lower slicing effectiveness.
In contrast, \ourtool attains an average latency of 0.296\,s and remains faster than other effective baselines.
Overall, \ourtool strikes a favorable trade-off between inference efficiency and slicing accuracy, demonstrating its practical viability.

\begin{table}[!htbp]
\centering
\scriptsize

\begin{tabular}{lcc}
\hline
\textbf{Methods} & \textbf{Size} & \textbf{Runtime (per task)}  \\
\hline
NS-slicer (CodeBERT) & 125M & 0.105s  \\
NS-slicer (GraphCodeBERT) & 125M & 0.135s  \\
\hline
CodeT5+ & 770M & 0.289s \\
\ourtool & 770M & 0.296s \\
\hline
CodeLlama-7B (SFT) & 7B & 5.75s \\
Qwen3-8B (SFT) & 8B & 6.52s  \\
\hline
GPT-4 (CoT) & Unknown & 0.835s \\
GPT-5 (CoT) & Unknown & 1.162s \\
\hline
\end{tabular}
\caption{Inference latency comparison among different program slicing methods on Java dataset.}
\label{tab:latency}
\vspace{-0.4cm}
\end{table}

%% file: rq3.tex


%% file: conclusion.tex
\section{Conclusion}\label{sec:conclusion}

This paper presents \ourtool, a learning-based approach for static program slicing that enhances accuracy through dataflow-aware pretraining and constrained decoding. We introduce two dataflow-aware pretraining objectives, statement permutation and span corruption, that enable the model to learn variable dependencies essential for precise slicing. In addition, we propose a constrained decoding strategy that enforces lexical and syntactic constraints, ensuring that generated slices are both accurate and faithful to the original program. We evaluate \ourtool on Java and Python programs from the CodeNet dataset, where it achieves state-of-the-art performance. Specifically, \ourtool outperforms the strongest baseline, NS-slicer, by 6.4\% and 21.9\% in ExactMatch on Java and Python, respectively.

%% file: limitation.tex



\section{Limitations}


\paragraph{Language limitations.}
We evaluate \ourtool on two widely used programming languages, Java and Python.
Although our approach is readily adaptable to other languages, additional engineering and empirical validation are required.
We encourage future work to extend our method to a broader range of programming languages and to other programming tasks whose outputs are subject to structural or semantic constraints.

\paragraph{Architecture limitations.}
Dataflow-aware pre-training is designed for encoder–decoder architectures, where masked-span reconstruction is naturally supported. In contrast, decoder-only models rely on next-token generation, making direct adoption of our pre-training objectives non-trivial. However, the proposed lexical–syntactic constrained decoding is architecture-agnostic and can be applied to both encoder–decoder and decoder-only models. Further discussion is provided in Appendix~\ref{app:architecture}.


\section{Ethical Considerations}
The implementation of this work is conducted with transparency, providing full disclosure of all technical details, limitations, and potential issues to the relevant stakeholders. The work avoids any false or misleading claims and ensures no data is fabricated or falsified.

In the interest of public benefit, the authors support reasonable and ethical uses of their intellectual contributions. Both the source code and data are released as free and open-source software and are made available in the public domain.

%% file: appendix.tex
\section{Statistics of the Two Language Datasets}\label{sec:qualitativeAnlysis}

\begin{table}[ht]
\caption{Statistics of the two language datasets}\label{tab:dataset}
\centering
\begin{tabular}{l|ccc}
\hline
\multicolumn{4}{c}{\textbf{Java}} \\ \hline
            & train    & valid    & test   \\ \hline
Entries     & 30.8K    & 3.5K     & 8.7K   \\
Avg. tokens & 64       & 64       & 66     \\
Avg. slocs  & 19       & 18       & 19     \\ \hline
\multicolumn{4}{c}{\textbf{Python}} \\ \hline
            & train    & valid    & test   \\ \hline
Entries     & 23.8K    & 5.1K     & 5.1K   \\
Avg. tokens & 88       & 86       & 91     \\
Avg. slocs  & 25       & 25       & 26     \\ \hline
\end{tabular}
\end{table}

Table~\ref{tab:dataset} reports the basic statistics of the Java and Python datasets. Both datasets are split into training, validation, and test sets. Java samples are generally shorter, with fewer tokens and source lines of code, while Python samples are relatively longer across all splits. SLOC (Source Lines of Code) counts the number of executable source lines, excluding blank lines and comments.

\section{Workflow of Constrained Decoding}\label{app:algorithms}

\begin{figure}[!htb]
  \centering
    {\includegraphics[width=1\linewidth]{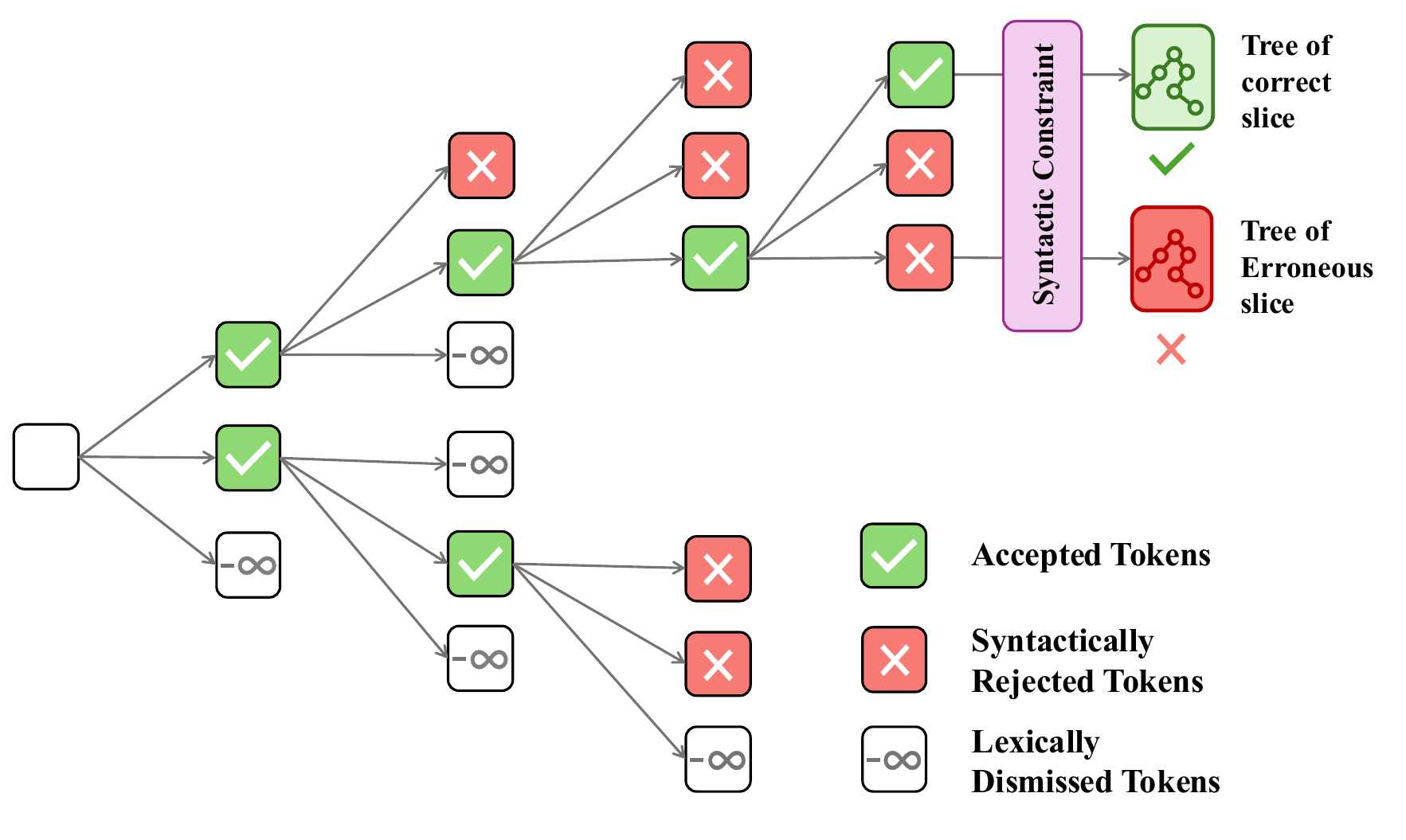}}
    \caption{Workflow of constrained decoding }\label{fig:decoding}
    \vspace{-0.3cm}
\end{figure}

Figure~\ref{fig:decoding} illustrates constrained decoding with a beam size of 3.
Each column presents the top three token predictions ranked by probability.
At each decoding step, lexical constraints filter out invalid tokens, while syntactic constraints based on TSED monotonicity prune malformed beams at statement boundaries.






\begin{table*}[!htbp]
\footnotesize
\centering
\small
\begin{tabular}{lccccc}
\hline
\textbf{Model} & \textbf{Architecture} & \textbf{Pretraining} & \textbf{Super Fine-tuning} & \textbf{Constrained Decoding} & \textbf{ExactMatch} \\
\hline
CodeT5        & Encoder--Decoder & $\times$ & \checkmark & $\times$ & 82.80 \\
CodeT5       & Encoder--Decoder & \checkmark & \checkmark & \checkmark & 85.12 \\
\hline
CodeLlama-7B  & Decoder-Only     & $\times$ & \checkmark & $\times$ & 75.27 \\
CodeLlama-7B  & Decoder-Only     & $\times$ & \checkmark & \checkmark & 78.40 \\
Qwen3-8B      & Decoder-Only     & $\times$ & \checkmark & $\times$ & 80.55 \\
Qwen3-8B      & Decoder-Only     & $\times$ & \checkmark & \checkmark & 83.11 \\
\hline
\end{tabular}
\caption{Comparison of transformer architectures and their compatibility with \ourtool components (Java).}
\label{tab:architecture} l.?
\end{table*}

\section{Qualitative Analysis}\label{sec:qualitativeAnlysis}

The main drawback of the previous SOTA, NS-slicer, lies in its design of task modeling.
For example, in Example (4) shown in Figure~\ref{fig:errornsslicer}, NS-slicer models program slicing as a binary classification task, with a fixed threshold to determine whether a statement belongs in the slice. As it embeds each statement independently, it fails to differentiate between identical statements at different positions. Consider the statement (e.g., \texttt{ch = true;}) appearing in two different branches of a method. NS-slicer treats both occurrences identically, even if only one is relevant. In contrast, \ourtool operates at the method level and leverages the full context, allowing it to accurately determine which occurrence is the accurate slice.
\begin{figure}[!htbp]
\begin{tcolorbox}
[boxsep=1pt,left=1pt,right=1pt,top=1pt,bottom=1pt]
\begin{lstlisting}[language=Java,frame=single,framerule=0pt]
// Example (4) - Erroneous slice from NS-slicer
// Expected slice:
...
19: if (Math.abs(as[l]) >= Math.abs(as[r])) {
20: lst.unset(r);
21: ch = true;
...
// Generated slice:
...
20: lst.unset(r);
21: ch = true;
22: }
23   (*@\textcolor{red}{ if (Math.abs(as[r]) >= Math.abs(as[l])) \{}@*)
24: (*@\textcolor{red}{lst.unset(l);}@*)
25: (*@\textcolor{red}{ch = true;}@*)
...
\end{lstlisting}
\end{tcolorbox}

\caption{An erroneous example predicted by NS-slicer.}
\label{fig:errornsslicer}
\end{figure}

\section{Transformer Architecture Comparison}\label{app:architecture}

We build \ourtool on top of the encoder--decoder model CodeT5+.
Prior work shows encoder--decoder architectures often achieve superior performance when trained on fixed datasets~\citep{encoder}.
To investigate whether our proposed techniques generalize across different transformer architectures, Table~\ref{tab:architecture} compares an additional encoder--decoder model (CodeT5) and two decoder-only models (CodeLlama and Qwen3) in terms of their compatibility with each component of \ourtool and their performance on the Java slicing task.

\textbf{Encoder--decoder models (CodeT5, CodeT5+).}
T5-style encoder--decoder models, including CodeT5+ (used in our experiments) and CodeT5, are naturally compatible with all components of \ourtool.
Their bidirectional encoder enables effective modeling of global code context, which is essential for both statement permutation and dataflow-aware span corruption during pre-training.
In particular, CodeT5 still achieves competitive results due to the inherent advantages of the encoder-decoder architecture for sequence-to-sequence program slicing.

\textbf{Decoder-only models (CodeLlama, Qwen3).}
Decoder-only models rely on causal attention and are pre-trained using next-token prediction under a strictly left-to-right generation paradigm.
As a result, they are fundamentally incompatible with our dataflow-aware pre-training objectives. However, they can still benefit from \textbf{constrained decoding} at inference time, which enforces lexical and syntactic constraints on generated outputs.